Data and text mining

# heatmaply: an R package for creating interactive cluster heatmaps for online publishing

Tal Galili[1,*], Alan O'Callaghan[2], Jonathan Sidi[3] and Carson Sievert[4]

[1]Department of Statistics and Operations Research, Tel Aviv University, Tel Aviv 6997801, Israel, [2]Fios Genomics, 9 Edinburgh Bioquarter, EH16 4UX, Scotland, UK, [3]Department of Statistics, Hebrew University, Jerusalem 9190401, Israel and [4]Department of Statistics, Iowa State University, 2438 Osborn Dr Ames, IA 50011-1090

*To whom correspondence should be addressed.
Associate Editor: Jonathan Wren

## Abstract

**Summary**: *heatmaply* is an R package for easily creating interactive cluster heatmaps that can be shared online as a stand-alone HTML file. Interactivity includes a tooltip display of values when hovering over cells, as well as the ability to zoom in to specific sections of the figure from the data matrix, the side dendrograms, or annotated labels. Thanks to the synergistic relationship between *heatmaply* and other R packages, the user is empowered by a refined control over the statistical and visual aspects of the heatmap layout.
**Availability and implementation**: The *heatmaply* package is available under the GPL-2 Open Source license. It comes with a detailed vignette, and is freely available from: http://cran.r-project.org/package=heatmaply.
**Contact**: tal.galili@math.tau.ac.il
**Supplementary information**: Supplementary data are available at *Bioinformatics* online.

## 1 Introduction

A cluster heatmap is a popular graphical method for visualizing high dimensional data. In it, a table of numbers is scaled and encoded as a tiled matrix of colored cells. The rows and columns of the matrix are ordered to highlight patterns and are often accompanied by dendrograms and extra columns of categorical annotation. The ongoing development of this iconic visualization, spanning over more than a century, has provided the foundation for one of the most widely used of all bioinformatics displays (Wilkinson and Friendly, 2009). When using the R language for statistical computing (R Core Team, 2016), there are many available packages for producing static heatmaps, such as: *stats*, *gplots*, *heatmap3*, *fheatmap* and *pheatmap* and others. Recently released packages also allow for more complex layouts; these include *gapmap*, *superheat* and *ComplexHeatmap* (Gu *et al.*, 2016). The next evolutionary step has been to create interactive cluster heatmaps, and several solutions are already available. However, these solutions, such as the *idendro* R package (Sieger *et al.*, 2017), are often focused on providing an interactive output that can be explored only on the researcher's personal computer. Some solutions do exist for creating shareable interactive heatmaps.

However, these are either dependent on a specific online provider, such as XCMS Online, or require JavaScript knowledge to operate, such as InCHlib. In practice, when publishing in academic journals, the reader is left with a static figure only (often in a png or pdf format).

To fill this gap, we have developed the *heatmaply* R package for easily creating a shareable HTML file that contains an interactive cluster heatmap. The interactivity is based on a client-side JavaScript code that is generated based on the user's data, after running the following command:

```
heatmaply(data, file = 'my_heatmap.html')
```

The HTML file contains a publication-ready, interactive figure that allows the user to zoom in as well as see values when hovering over the cells. This self-contained HTML file can be made available to interested readers by uploading it to the researcher's homepage or as a Supplementary Material in the journal's server. Concurrently, this interactive figure can be displayed in RStudio's viewer pane, included in a Shiny application, or embedded in a knitr/RMarkdown HTML documents.

The rest of this paper offers guidelines for creating effective cluster heatmap visualization. Figure 1 demonstrates the suggestions from this section on data from project Tycho (van Panhuis *et al.*, 2013), while the online Supplementary Material includes the interactive version, as well as several examples of using the package on real-world biological data.

## 2 heatmaply: a simple example

The generation of cluster heatmaps is a subtle process (Gehlenborg and Wong, 2012; Weinstein, 2008), requiring the user to make many decisions along the way. The major decisions to be made deal with the data matrix and the dendrogram. The raw data often need to be transformed in order to have a meaningful and comparable scale, while an appropriate color palette should be picked. The clustering of the data requires us to decide on a distance measure between the observation, a linkage function, as well as a rotation and coloring of branches that manage to highlight interpretable clusters. Each such decision can have consequences on the patterns and interpretations that emerge. In this section, we go through some of the arguments in the function heatmaply, aiming to make it easy for the user to tune these important statistical and visual parameters. Our toy example visualizes the effect of vaccines on measles infection. The output is given in the static Figure 1, while an interactive version is available online in the Supplementary file 'measles.html'. Both were created using:

```
heatmaply(x = sqrt(measles),
    color = viridis, # the default
    Colv = NULL,
    hclust_method = 'average', k_row = NA, # ...
    file = c('measles.html', 'measles.png'))
```

The first argument of the function (x) accepts a matrix of the data. In the measles data, each row corresponds with a state, each column with a year (from 1928 to 2003), and each cell with the number of people infected with measles per 100 000 people. In this example, the data were scaled twice—first by not giving the raw number of cases with measles, but scaling them relatively to 100 000 people, thus making it possible to more easily compare between states. And second by taking the square root of the values. This was done since all the values in the data represent the same unit of measure, but come from a right-tailed distribution of count data with some extreme observations. Taking the square root helps with bringing extreme observations closer to one another, helping to avoid an extreme observation from masking the general pattern. Other transformations that may be considered come from Box-Cox or Yeo-Johnson family of power transformations. If each column of the data were to represent a different unit of measure, then leaving the values unchanged will often result in the entire figure being un-usable due to the column with the largest range of values taking over most of the colors in the figure. Possible per-column transformations include the scale function, suitable for data that are relatively normal. normalize, and percentize functions bring data to the comparable 0–1 scale for each column. The normalize function preserves the shape of each column's distribution by subtracting the minimum and dividing by the maximum of all observations for each column. The percentize function is similar to ranking but with the simpler interpretation of each value being replaced by the percent of observations that have that value or below. It uses the empirical cumulative distribution function of each variable on its own values. The sparseness of the dataset can be explored using is.na10.

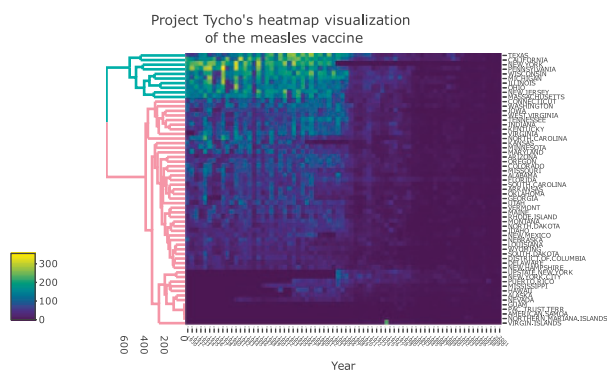

**Fig. 1.** The (square root) number of people infected by Measles in 50 states, from 1928 to 2003. Vaccines were introduced in 1963, An interactive version is available in the following URL: https://cdn.rawgit.com/talgalili/heatmaplyExamples/564da09e/inst/doc/measles_heatmaply.html

Once the data are adequately scaled, it is important to choose a good color palette for the data. Other than being pretty, an ideal color palette should have three (somewhat conflicting) properties: (i) Colorful, spanning as wide a palette as possible so as to make differences easy to see; (ii) Perceptually uniform, so that values close to each other have similar-appearing colors compared with values that are far away, consistently across the range of values; and (iii) Robust to colorblindness, so that the above properties hold true for people with common forms of colorblindness, as well as printing well in grey scale. The default passed to the color argument in heatmaply is viridis, which offers a sequential color palette, offering a good balance of these properties. Divergent color scale should be preferred when visualizing a correlation matrix, as it is important to make the low and high ends of the range visually distinct. A helpful divergent palette available in the package is cool_warm (other alternatives in the package include RdBu, BrBG, or RdYlBu, based on the *RColorBrewer* package). It is also advisable to set the limits argument to range from -1 to 1.

Passing NULL to the Colv argument, in our example, removed the column dendrogram (since we wish to keep the order of the columns, relating to the years). The row dendrogram is automatically calculated using hclust with a Euclidean distance measure and the average linkage function. The user can choose to use an alternative clustering function (hclustfun), distance measure (dist_method), or linkage function (hclust_method), or to have a dendrogram only in the rows/columns or none at all (through the dendrogram argument). Also, the users can supply their own dendrogram objects into the Rowv (or Colv) arguments. The preparation of the dendrograms can be made easier using the *dendextend* R package (Galili, 2015) for comparing and adjusting dendrograms. These choices are all left for the user to decide. Setting the k_col/k_row argument to NA makes the function search for the number of clusters (from 2 to 10) by which to color the branches of the dendrogram. The number picked is the one that yields the highest average silhouette coefficient (based on the find_k function from *dendextend*). Lastly, the heatmaply function uses the *seriation* package to find an 'optimal' ordering of rows and columns (Hahsler *et al.*, 2008). This is controlled using the seriation argument where the default is 'OLO' (optimal-leaf-order)—which rotates the branches so that the sum of distances between each adjacent leaf (label) will be minimized (i.e.: optimize the Hamiltonian path length that is restricted by the dendrogram structure). The other arguments in the example were omitted since they are self-explanatory—the exact code is available in the heatmaplyExamples package.

In order to make some of the above easier, we created the shinyHeatmaply package (available on CRAN) which offers a GUI

to help guide the researcher with the heatmap construction, with the functionality to export the heatmap as an html file and summaries parameter specifications to reproduce the heatmap with heatmaply. For a more detailed step-by-step demonstration of using heatmaply on biological datasets, you should explore the heatmaplyExamples package (https://github.com/talgalili//heatmaplyExamples).

## Acknowledgements

The heatmaply package was made possible by leveraging many wonderful R packages, including *ggplot2* (Wickham, 2009), *plotly* (Sievert *et al.*, 2016), *dendextend* (Galili, 2015) and many others. We would also like to thank Yoav Benjamini, Madeline Bauer and Marilyn Friedes for their helpful comments, as well as Joe Cheng for initiating the collaboration with Tal Galili on d3heatmap, which helped lay the foundation for heatmaply.

## Funding

This work was supported in part by the European Union Seventh Framework Programme (FP7/2007–2013) under grant agreement no. 604102 (Human Brain Project).

*Conflict of Interest*: none declared.